\documentclass[11pt]{gh2013}

\usepackage{mathptmx}       % selects Times Roman as basic font
\usepackage{helvet}         % selects Helvetica as sans-serif font
\usepackage{courier}        % selects Courier as typewriter font
\usepackage{makeidx}         % allows index generation
\usepackage{graphicx}        % standard LaTeX graphics tool
\usepackage{multicol}        % used for the two-column index
\begin{document}

\title*{The Processing of the Clumpy Molecular Gas in the Galactic Center and the Star-Formation}
\titlerunning{Star-Formation in the Sgr\,A Molecular Clouds}
\author{Hauyu Baobab Liu$^{1}$, Paul T.~P. Ho$^{1,2}$, Melvyn C. H. Wright$^{3}$, Yu-Nung Su$^{1}$, Pei-Ying Hsieh$^{1,4}$, Ai-Lei Sun$^{5}$, Sungsoo S. Kim$^{6}$, and Young Chol Minh$^{7}$}
\authorrunning{Liu, Hauyu Baobab et al.} 
\institute{$^{1}$Academia Sinica Institute of Astronomy and Astrophysics (ASIAA), \email{hyliu@asiaa.sinica.edu.tw}
\\ $^{2}$ Harvard-Smithsonian Center for Astrophysics
\\ $^{3}$ Radio Astronomy Laboratory, University of California, Berkeley
\\ $^{4}$ Graduate Institute of Astronomy, National Central University
\\ $^{5}$ Department of Astrophysical Sciences, Peyton Hall, Princeton University
\\ $^{6}$ Department of Astronomy and  Space Science, Kyung Hee University
\\ $^{7}$ Korea Astronomy and Space Science Institute (KASI)
 }
\maketitle

%\abstract*{}

\vskip -3 cm  %%% Please Modify that Value if more than one Line of Authors 
\abstract{
We present the Green Bank 100\,m Telescope (GBT) mapping observations of CS 1-0, and the Submillimeter Array (SMA) 157-pointings mosaic of the 0.86 mm dust continuum emission as well as several warm and dense gas tracers, in the central $\sim$20 pc area in Galactic Center. 
The unprecedentedly large field-of-view and the high angular resolution of our SMA dust image allow the identification of abundant 0.1-0.2 pc scale dense gas clumps. We found that in the Galactic Center, the Class I methanol masers are excellently correlated with the dense gas clumps. 
However, on the ~0.1 pc scale, these dense gas clumps still have a extremely large linewidth (FWHM$\sim$10-20 km\,s$^{-1}$). 
Simple calculations suggest that the identified clumps can be possibly the pressurized gas reservoir feeding the formation of 1-10 solar-mass stars. 
These gas clumps may be the most promising candidates for ALMA to resolve the high-mass star-formation in the Galactic center.
} 

\section{Introduction}
\label{sec:back}
The universality of the star formation laws has been widely investigated in extragalactic studies ([1], and references therein).
When taking a closer look, these star formation laws may represent merely the weighted media for the large number of observed molecular clouds.
Ultimately, the high precision observations may allow us to parametrize the details about how the star formation rate and efficiency are deviated from the laws, because of the different physical conditions in the natal molecular clouds. 
Preliminary hints may first be provided by observing the environments with extreme physical conditions. 

The $\sim$200 pc Galactic Central Molecular Zone (CMZ) provides the unique laboratory to resolve the star-forming molecular clouds around the supermassive black hole with the sub-parsec scale resolutions, thanks to the proximity.
The high molecular gas temperature and the largely non-virialized gas motions in these clouds present the most extreme environment for star formation ([2], and references therein). 
The previous observations on molecular cloud M0.25+0.01 in the CMZ have reported the $\sim$2 orders of magnitude smaller star-forming efficiency than the molecular clouds in the rest of the Galactic disk [3, 4, 5, 6].
In this work, we present the multi-scale mapping observations towards the Sgr\,A molecular clouds, which are the star-forming regions nearest to the central supermassive black hole Sgr\,A*. 
Our aims were to resolve the overall geometry and kinematics of these molecular clouds, and to examine the formation of the localized gas overdensities in high resolution images. 
Our large-scale mapping observations have provided the road maps for the higher angular resolution follow-ups. 

\vspace{-0.3cm}
\begin{figure}[h]
\hspace{-0.9cm}
\includegraphics[width=12.5cm]{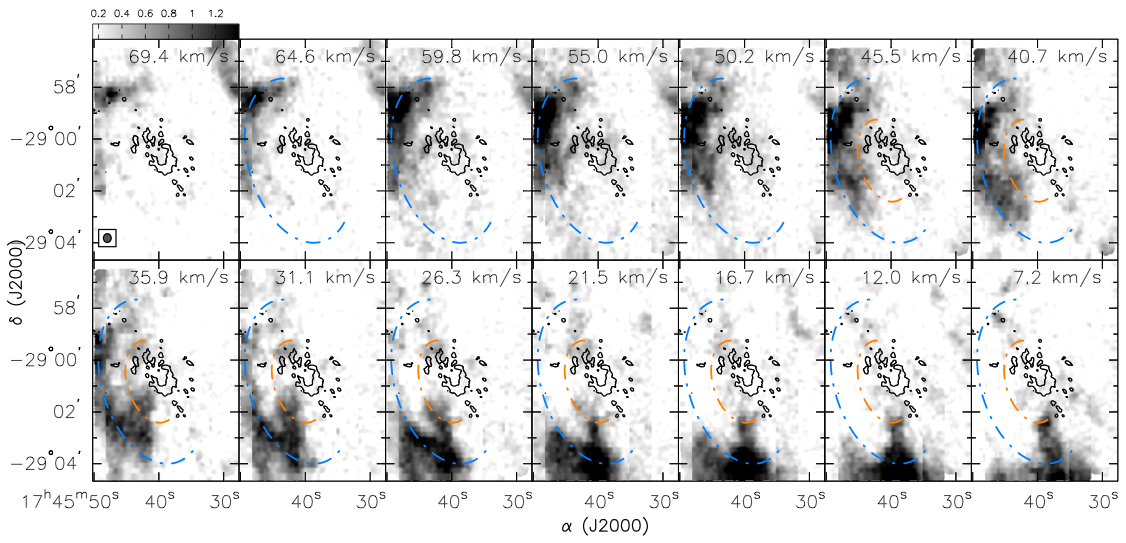}
\vspace{-0.5cm}
\caption{\small{
The channel maps of the CS 1--0 line (gray scale; [7]). The color bar is in Kelvin units. White contours show the 300\,Jy\,beam$^{-1}$\,km\,s$^{-1}$ level of the velocity integrated HCN 4--3 intensity map taken by SMA, to indicate the warm CND. The orange dashed-dotted arc is drawn to indicate the newly discovered Southern Arc. The light blue dashed-dotted arc is drawn to indicate the Eastern Arm constituted by the well-known 50 km\,s$^{-1}$ molecular cloud, the northern part of the 20 km\,s$^{-1}$ molecular cloud, and the molecular ridge located in between.
}}
\label{fig_arc}
\vspace{-0.8cm}
\end{figure}

\section{Observations}
\label{sec:obs}
We observed the CS 1-0 transition and the NH$_{3}$ from (1,1) to (6,6) transitions using the NRAO GBT [7, 8]. 
We made the 157-pointings mosaic observations (354.1-358.1 GHz and 342.1-346.1 GHz) toward the Galactic center using the SMA, in its compact and subcompact array configurations [7, 9].
The minimum and maximum projected baselines in our SMA observations are $\sim$7.0 $k\lambda$ and $\sim$82 $k\lambda$.  
The short spacing continuum data were complemented by combining the archival JCMT SCUBA image, using the procedure outlined in our previous papers [9, 10].
The zeroth order free-free continuum emission model was constructed based on the archival VLA 7 mm observation data, and was subtracted from the combined SMA+JCMT 0.86 mm continuum image.

\section{Results}
\label{sec:res}
\begin{figure}[h]
\vspace{-1.5cm}
\hspace{0cm}
\includegraphics[width=12cm]{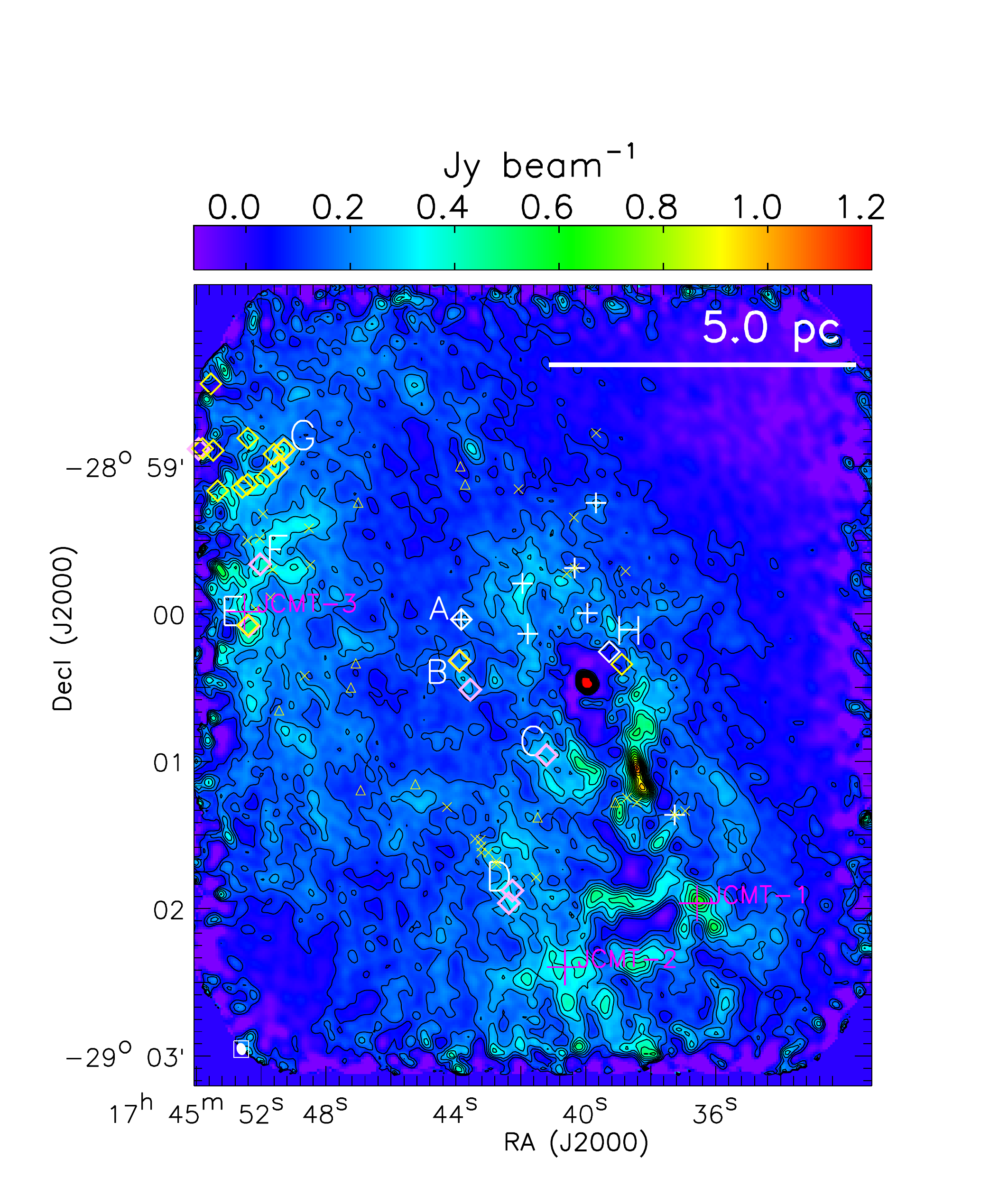} 
\vspace{-0.9cm}
\caption{\footnotesize{
The SMA+JCMT 0.86 mm continuum image with a free-free model subtracted. (color and contour; [9]). 
Contour spacings are 3$\sigma$ starting at 3$\sigma$ ($\sigma$=24 mJy\,beam$^{-1}$).
Yellow crosses are the 1720 MHz OH masers [14].
Yellow triangles are the compact, either thermal or low-gain masing 1612 MHz OH line sources [15].
Pink diamonds are the 36.2 GHz Class I CH$_{3}$OH masers reported in [16].
Yellow diamonds are the 44.1 GHz Class I CH$_{3}$OH masers reported in [17].
White Crosses and diamonds are the 22 GHz water masers and the 44.1 GHz CH$_{3}$OH masers reported in [18].
}}
\vspace{-0.3cm}
\label{fig_ch0ffsub}
\end{figure}

Figures \ref{fig_arc} and \ref{fig_ch0ffsub} show the CS 1-0 velocity channel maps, and the SMA+JCMT 0.86 mm dust image [7, 9]. 
These observations covered the well-known 50 km\,s$^{-1}$ molecular cloud, the northern part of the 20 km\,s$^{-1}$ molecular cloud, and the 2-4 pc scale circumnuclear disk (CND) surrounding the central supermassive black hole Sgr\,A*.
The CS 1-0 channel maps show that the 50 km\,s$^{-1}$ and the northern part of the 20 km\,s$^{-1}$ molecular clouds are connected by the molecular ridge in the east. 
They are likely to be sectors of the same dynamical entity. 
Since the early careful analysis of the formaldehyde absorption line have suggested that the 50 km\,s$^{-1}$ and the 20 km\,s$^{-1}$ molecular clouds are background and foreground to the Sgr\,A*, respectively [11]. 
We proposed that these gas clouds (as well as the newly discovered Southern Arc; Figure \ref{fig_arc}) are the gravitationally accelerated gas streams orbiting around the Sgr\,A* and the central stellar cluster at the small radii (e.g. $\ge$5-10 pc) [7]. 
The asymmetric and clumpy CND may be the tidally trapped inner part of these gas streams.
We note that the entire 20 km\,s$^{-1}$ molecular cloud consists of multiple components in the line of sight [8]; other components may locate at the further distances from the Sgr\,A* [12].

These molecular clouds/streams are subjected to strong ionizing flux and the pressure of the hot ionized gas, however, still show high-mass star-forming activities including the ultracompact H\textsc{ii} regions in the west of the 50 km\,s$^{-1}$ cloud [13]. 
Our dust continuum image marginally resolves at least several tens of 10-10$^{2}$ $M_{\odot}$ dense clumps in the 5$'$ field, including the CND and the exterior gas streamers (Figure \ref{fig_ch0ffsub}), which provides the candidates of gas reservoirs to feed the massive star formation.  
From the spectral line data, the velocity dispersions of the dense clumps and their parent molecular clouds are $\sim$10-20 times higher than their virial velocity dispersions [9], which is contrast to the typically virialized gas cores in high-mass star-forming regions outside of the CMZ [14]. 
Some of the dense gas clumps are associated with 22 GHz water masers and 36.2 GHz and 44.1 GHz CH$_{3}$OH masers (Figure \ref{fig_ch0ffsub}). 
However, we do not find clumps which are bound by the gravity of the enclosed molecular gas.
The detailed kinematics in these gas clumps remains to be resolved in higher resolution observations.

\begin{acknowledgement}
HBL sincerely acknowledge the organizers of the Guillermo Haro conference where these results were presented. HBL thanks Dr. Luis F. Rodr\'{i}guez for the supports. 
\vspace{-0.3cm}
\end{acknowledgement}

\end{document}